\documentclass[fleqn,usenatbib]{mnras}

\usepackage{newtxtext,newtxmath}
\usepackage[T1]{fontenc}

\DeclareRobustCommand{\VAN}[3]{#2}
\let\VANthebibliography\thebibliography
\def\thebibliography{\DeclareRobustCommand{\VAN}[3]{##3}\VANthebibliography}

\usepackage{graphicx}	% Including figure files
\usepackage{amsmath}	% Advanced maths commands

\usepackage{amssymb}	% Extra maths symbols
\usepackage{longtable}  % long table
\usepackage{subfigure}  % subfigure
\usepackage{caption}
\usepackage{multirow}
\usepackage{lastpage}

\newcommand{\kms}{\ensuremath{\mathrm{km}\,\mathrm{s}^{-1}}}

 \newcommand{\LCDM}{$\Lambda$CDM}

\newcommand{\Msun}{\ensuremath{\rm{{M}_{\odot}}}}

\newcommand{\hi}{H{\sc i} }

\usepackage{graphicx}	% Including figure files
\usepackage{epstopdf}

\usepackage{amsmath}	% Advanced maths commands
\usepackage{amssymb}	% Extra maths symbols

\usepackage{lineno}

\usepackage{pdflscape}
\usepackage{orcidlink}

\title[Extraction of \hi gas with bulk motions in the disk of galaxies]{Extraction of \hi gas with bulk motions in the disk of galaxies}

\author[S-H. Oh et al.]{
Se-Heon Oh$^{1}$\thanks{E-mail: seheon.oh@sejong.ac.kr}
and Jing Wang$^{2}$\thanks{E-mail: jwang\_astro@pku.edu.cn}
\\
$^{1}$Department of Physics and Astronomy, Sejong University, 209 Neungdong-ro, Gwangjin-gu, Seoul, Republic of Korea \\
$^{2}$Kavli Institute for Astronomy and Astrophysics, Peking University, Beijing 100871, China}

\date{Accepted for publication in MNRAS}

\pubyear{2022}

\begin{document}
\label{firstpage}
\pagerange{\pageref{firstpage}--\pageref{lastpage}}
\maketitle

\begin{abstract}
We propose a new method for extracting bulk motion gases in the disk of a galaxy from \hi data cubes, offering improvements over classical techniques like moment analysis and line profile fitting. Our approach decomposes the line-of-sight velocity profiles into multiple Gaussian components, which are then classified into (underlying and dominant) bulk and non-bulk motion gases based on criteria such as \hi surface density, velocity dispersion, kinetic energy, and rotation velocity. A 2D tilted-ring analysis is employed to refine the kinematical parametres of the galaxy disk, ensuring robust extraction of the bulk motion gases. We demonstrate the effectiveness of this method using the \hi data cubes of NGC 4559 from the WSRT-HALOGAS survey, distinguishing between bulk and non-bulk gas components. From this, we find that approximately 50\% of the \hi gas in NGC 4559 is classified as non-bulk, possibly linked to processes such as stellar feedback. This work provides a robust framework for analysing \hi kinematics of galaxies from high sensitivity \hi observations of galaxies like MeerKAT-MHONGOOSE and FAST-FEASTS and allows us to best exploit the kinematic information of the complex gas dynamics within galaxy disks.

\end{abstract}


%%%%%%%%%%%%%%%%%%%%%%%%%%%%%%%%%%%%%%%%%%%%%%%%%%

%%%%%%%%%%%%%%%%% BODY OF PAPER %%%%%%%%%%%%%%%%%%

\section{Introduction}\label{sec:1}

The dynamics of gas in a galaxy's disk are governed by the galaxy's total gravitational potential, which arises from the combined contributions of baryonic matter (such as stars and gas) and the dark matter halo. The dark matter halo typically accounts for more than 50\% and up to 90\% of the galaxy's mass, depending on the galaxy type, within approximately five effective radii of the stellar disk, (e.g., \citealt{2020ApJ...905...28H, 2021A&A...653A..20S}).

\hi gas, being the dominant gaseous component in the disk, is usually considered to have a thin vertical distribution and extends radially to about three times the radius of the stellar disk \citep{2007A&A...469..511K, 2009ARA&A..47...27K}. In spiral galaxies, where circular rotation dominates over non-circular motions, the \hi gas emitting the 21 cm hyperfine transition line is an effective tracer of global kinematics. By extracting the line-of-sight velocities of individual \hi gas clouds across the disk and compiling them into a 2D map—the so-called \hi velocity field—one can derive the galaxy’s rotation curve.

This velocity field, under the assumption of an infinitely thin gas disk, can be fitted with various 2D kinematic models. These include theoretical models like the \LCDM\, dark matter halo model (NFW: \citealt{1996ApJ...462..563N}), observationally motivated models such as the pseudo-isothermal halo model (ISO: \citealt{1987gady.book.....B}), and empirically derived functional forms like an arctan \citep{2008A&A...484..173P}, an inverted exponential \citep{2011RAA....11.1429F}, or a hyperbolic tanh \citep{2013ApJ...768...41A} rotation curve models. Non-parametric kinematic models, such as the widely adopted tilted-ring analysis using concentric tilted rings (\citealt{1989A&A...223...47B}), are also commonly used for analysing the kinematics of galaxies, particularly using \hi observations. While variations of the tilted-ring approach may involve different levels of parameterisation or constraints, the fundamental methodology remains consistent across these implementations.

When utilizing the 2D line-of-sight velocity field extracted from the \hi data cube, the accuracy of the kinematic analysis is critically dependent on the extracted velocity field. Several methods have been employed to determine the representative central velocity of a given line-of-sight velocity profile: (1) intensity-weighted mean velocity (moment1, e.g.,  \citealt{2013pss5.book..985S}), (2) single Gaussian fitting (e.g., \citealt{2007IAUS..235..193C, 2022ApJ...928..177O}), (3) Hermite $h3$ Gaussian fitting (e.g., \citealt{2008AJ....136.2648D}), and (4) multiple-Gaussian fitting (e.g., \citealt{2011AJ....141..193O, 2015AJ....149..180O}). For a gas disk with an intermediate inclination (e.g., 40–70 degrees; \citealt{2008AJ....136.2648D}) and a purely circular rotation velocity field, all these methods can determine the most representative central velocities of the gas profiles. However, as non-circular motions become more significant, the differences among the derived central velocities using these methods increase, deviating from the bulk circular rotation \citep{2018MNRAS.473.3256O}.

\citet{2019MNRAS.485.5021O} introduced a new approach to extract the circularly rotating gas motions in the disk, referred to as the bulk velocity field, based on the multiple Gaussian decomposition of line-of-sight velocity profiles from an \hi data cube. This method involves decomposing the individual \hi velocity profiles into an optimal number of Gaussian components using Bayesian statistics and classifying these components into circularly rotating components (bulk motions) and non-circular motions based on a reference model velocity field. The reference velocity field can be generated from the \hi data cube using methods like single Gaussian fitting or {\sc moment1}, and can be further refined with the bulk velocity field obtained from the initial iteration. Not only H{\sc i}, but also other kinematic tracers such as CO, H$\alpha$ emission, or stellar absorption lines, can be used to trace the bulk rotation of galaxies when constructing the model reference velocity field (e.g. \citealt{2022ApJ...928..177O}). This iterative approach helps correct for the effects of non-circular motions, allowing for the extraction of the bulk motion gas in the disk, which are then used to derive the global bulk rotation curve of a galaxy for fitting 2D kinematic models. However, when non-circular gas motions dominate over circular motions in the disk, the aforementioned methods, which rely on kinematic information during classification, may fail to accurately classify the decomposed Gaussian components and, consequently, the resulting bulk velocity field.

From a galaxy simulation perspective, the distribution and motion of individual \hi gas clouds within the disk are significantly influenced by gravitational and hydrostatic forces during the galaxy’s formation and evolution, resulting in characteristic features. Firstly, the integrated flux of \hi gas clouds decreases with galactic radius (\citealt{2016MNRAS.456.1115B}). Observationally, a central depression in \hi flux is often seen in the central regions of galaxies, primarily caused by either the consumption of \hi gas into stars or its transition into $\rm H_{2}$ molecules, which is described by a double exponential disk model \citep{2014MNRAS.441.2159W}. Secondly, in most galaxies, the velocity dispersion of \hi gas clouds decreases exponentially with galactic radius \citep{2009MNRAS.392..294A}. \citet{2009AJ....137.4424T} reported that their sample exhibits a characteristic velocity dispersion of \(10 \pm 2\) km s\(^{-1}\) at \(R_{25}\). By comparing the predicted energy input from various physical mechanisms with the kinetic energy of the gas inferred from the observed linewidths, they concluded that star formation is the dominant mechanism determining the profile widths within \(R_{25}\). Outside this radius, ultraviolet (UV) heating from extragalactic sources or magneto-rotational instability (MRI) are likely candidates influencing the profile widths. Thirdly, the kinetic energy, defined as 1.5 times the \hi surface density multiplied by the square of the velocity dispersion per unit area in the \hi gas within a galaxy’s disk, also decreases exponentially. Several energy sources can drive the ISM turbulence, including (1) star formation processes (stellar winds, UV ionization, supernova explosions; e.g., \citealt{2004A&A...425..899D, 2013ApJ...776....1K}), (2) galaxy kinematics (rotational shear, non-linear gravitational instability, spiral arms; e.g., \citealt{2004ApJ...609..667S, 2007ApJ...660.1232K, 2022MNRAS.509.2979N}), (3) magnetized disk instability (e.g., \citealt{2004MNRAS.349..270W, 2006ApJ...646..213K}), and (4) thermal instability (e.g., \citealt{2006astro.ph.12779H}).

These characteristic properties of the \hi 21 cm line in the disk of a galaxy—surface density, velocity dispersion, and kinetic energy—can complement the identification and extraction of bulk-disk-gas components when combined with the kinematic analysis described above. In this paper, we introduce an improved method for extracting global/bulk motion gas in the disk of a galaxy from \hi data cubes. Specifically, we improve the previous bulk velocity field extraction method described by \citet{2019MNRAS.485.5021O} by incorporating the characteristic \hi 21 cm line properties during the classification of the decomposed Gaussian components across the disk, alongside a 2D rotation curve analysis. We demonstrate this method by applying it to the \hi data cubes of NGC 4559 from the WSRT\footnote{Westerbork Synthesis Radio Telescope}\--HALOGAS (\citealt{2011A&A...526A.118H}) survey. 

The structure of the remaining sections of this paper is as follows: Section~\ref{sec:2} describes the procedure of the method for extracting bulk motion gas in the disk of a galaxy from an \hi data cube. For this, we use the \hi data cube of NGC 4559 from the WSRT-HALOGAS survey. In Section~\ref{sec:3}, we derive the bulk rotation curve of NGC 4559 using the extracted bulk velocity field. Section~\ref{sec:4} discusses non-bulk \hi gas motions in the disk of NGC 4559 which deviate from the bulk kinematics in the position-velocity and phase-space diagrams. Lastly, Section~\ref{sec:5} provides a summary of the main findings of this study.

\section{Procedure for extracting bulk motion gas in the disk of a galaxy from an \hi data cube: The case of NGC 4559}\label{sec:2}

\begin{figure*}
    \includegraphics[width=1.0\textwidth]{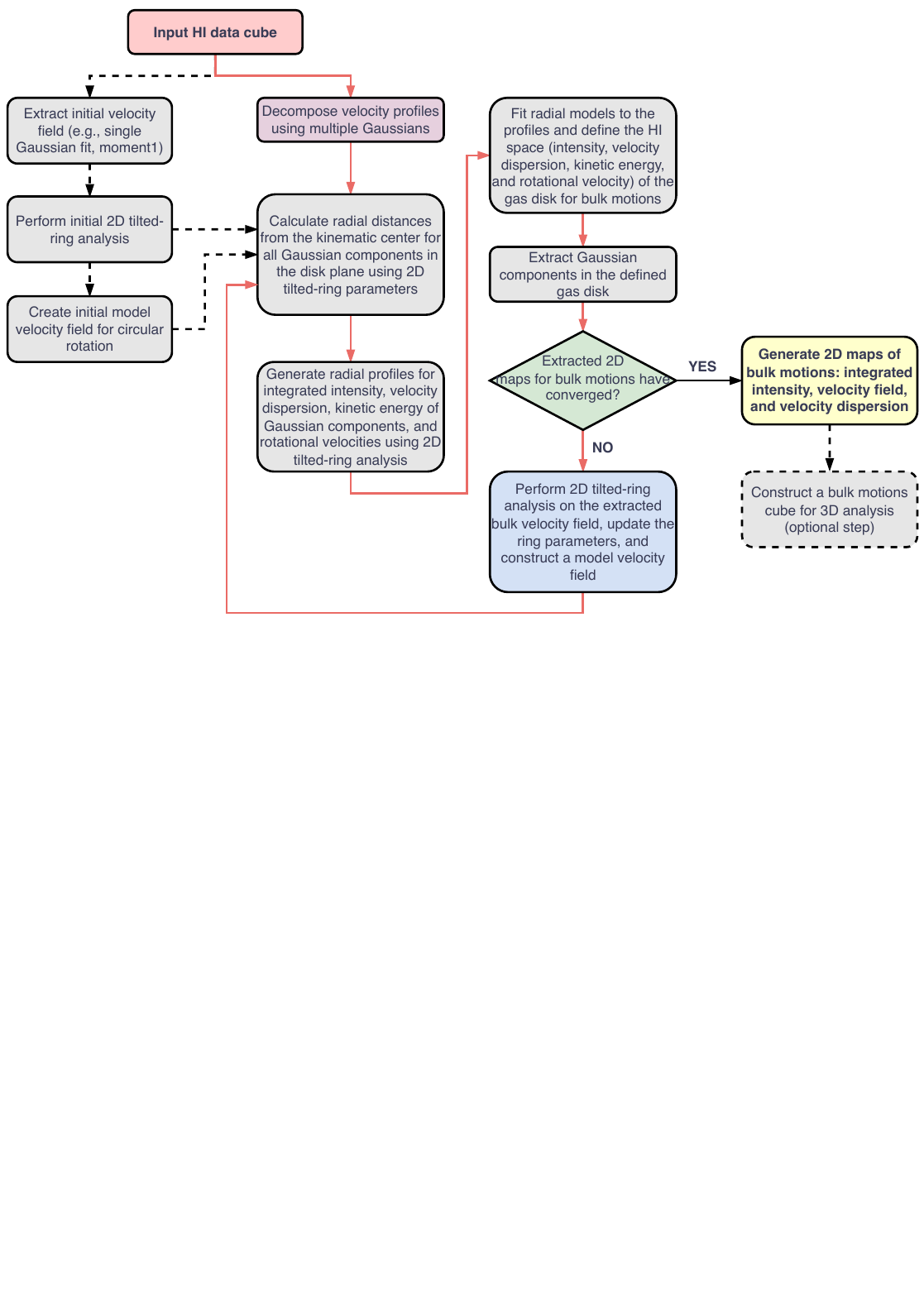}
    \vspace{-130mm} % Adjust the value as needed
    \caption{Schematic of the bulk motion gas extraction method from an \hi data cube. The process involves extracting the initial velocity field, applying a 2D tilted-ring analysis, decomposing profiles into multiple Gaussians, and classifying components into bulk and non-bulk motions based on radial models of intensity, velocity dispersion, and kinetic energy. Iterative refinement is performed until convergence, with an optional step to construct a bulk motions cube for further 3D analysis.}
    \label{fig:1}

\end{figure*}

In this section, we outline the schematic procedure for extracting circularly rotating bulk motion gas in the disk of a galaxy from an \hi data cube. This process is typically applied to data cubes obtained from radio interferometry, formatted with sky position coordinates (RA, Dec) along the line-of-sight velocity axis (velocity). As illustrated in Fig.~\ref{fig:1}, the procedure involves identifying and classifying gas motions into bulk (circularly rotating) and non-bulk components which deviate from the bulk motions. We demonstrate this step-by-step process using the high-resolution \hi data cube of NGC 4559 (natural-weighted, \( 28\arcsec.4 \times 13\arcsec.1\) spatial resolution, 4.12 \kms\ spectral resolution) taken from the WSRT-HALOGAS survey (\citealt{2011A&A...526A.118H}).

\begin{figure*}
    \hspace{-10mm}
    \includegraphics[width=1.0\textwidth]{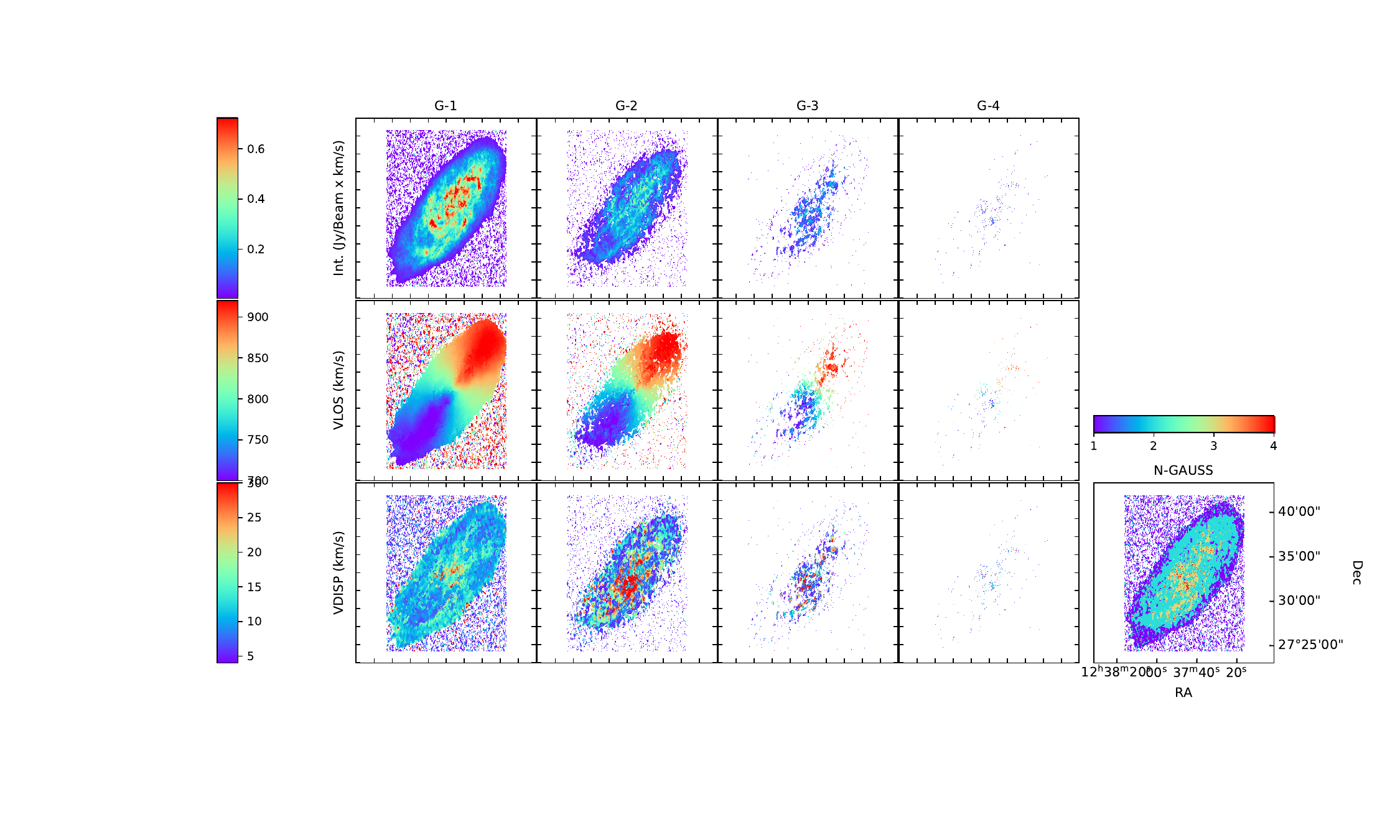}
    \vspace{-20mm} % Adjust the value as needed
    \caption{2D maps of the Gaussian components extracted from the \hi data cube of NGC 4559, as described in Section~\ref{sec:2.1}. For display purposes, up to four Gaussian components (G1, G2, G3, and G4) are shown: (1) integrated intensity (Int.) in units of $\rm Jy\,beam^{-1}\,km\,s^{-1}$, (2) line-of-sight velocity (VLOS) in units of $\rm km\,s^{-1}$, and (3) velocity dispersion (VDISP) in units of $\rm km\,s^{-1}$. G1, G2, G3, and G4 are sorted by their integrated intensity.}
    \label{fig:2}
\end{figure*}

\subsection{Profile decomposition}\label{sec:2.1}

The first step in our method is to decompose all line-of-sight velocity profiles in the \hi data cube into an optimal number of Gaussian components by fitting multiple Gaussian models to each profile. The objective is to extract individual gas clouds that adhere to the global \hi properties of the disk, such as intensity, velocity dispersion, kinetic energy, and kinematics, as discussed in Section~\ref{sec:1}. Conceptually, we treat the disk as a bulky system composed of elementary particles (i.e., individual \hi gas clouds) that are axisymmetrically distributed under the galaxy's total gravitational potential, and described by models for gas radial profiles like exponential or double-exponential fits.

Gas clouds that deviate from the global \hi properties of the disk can be isolated through the profile decomposition of the velocity profiles. For this purpose, we employ the {\sc baygaud-PI}\footnote{https://github.com/seheon-oh/baygaud-PI} tool, which decomposes a spectral line profile into multiple Gaussian components using Bayesian nested sampling. A distinctive feature of this tool is its ability to determine the optimal number of Gaussian components for a given line profile based on Bayes factor statistics. This facilitates the extraction of various Gaussian 2D maps from the line-of-sight velocity profiles in an \hi data cube, which can then be utilized in subsequent \hi kinematic analyses, such as deriving the galaxy rotation curve. A comprehensive description of the algorithm and its practical applications to \hi data cubes can be found in \citet{2019MNRAS.485.5021O, 2022ApJ...928..177O, 2023MNRAS.519..318K}, and \citet{2023ApJ...944..102W}.

Fig.~\ref{fig:2} illustrates the profile decomposition results for the HALOGAS \hi data cube of NGC 4559, derived using {\sc baygaud-PI}. As discussed in \citet{2022ApJ...928..177O}, when running {\sc baygaud-PI}, it is necessary to set the maximum number of Gaussian components (N-max-Gauss) to which the multiple Gaussian components will be fitted. The velocity profiles of the cube can be visually inspected prior to setting this parametre. For NGC 4559, we set N-max-Gauss to 6, which is sufficiently high to model even the most extreme cases of non-Gaussian profiles in the data cube. In this case, {\sc baygaud-PI} fits a series of models with multiple Gaussian components, ranging from 1 to 6, to each \hi velocity profile and selects the most appropriate Gaussian model using Bayes factor-based model selection. We adopt a strong Bayes factor criterion of 100 between any two competing models. The resulting optimal number of Gaussian components for all the lines of sight is mapped in bottom-rightmost panel of Fig.~\ref{fig:2}. It is noteworthy that the signal-to-noise (S/N) ratios of all the optimally decomposed Gaussian components are greater than 3. As shown in the figure, parts of the disk are described by triple or even quadruple Gaussian components, although the majority is described by double Gaussian components. The central region may be affected by observational projection effects, but could also be influenced by hydrodynamical processes, such as star formation or gravitational effects in the galaxy.

The integrated intensity, central velocity, and velocity dispersion of the decomposed Gaussian components are mapped in panels of Fig.~\ref{fig:2}, up to the maximum number, four of components in the N-Gauss map shown in the bottom-rightmost panel. For clarity, the Gaussian components are ordered according to their integrated intensities from G-1 to -3. In the following sections, we outline the selection criteria used to distinguish between bulk and non-bulk motion gases from these decomposed Gaussian components.

\subsection{Initial 2D tilted-ring analysis}\label{sec:2.2}

\begin{figure*}
    \includegraphics[width=1.0 \textwidth]{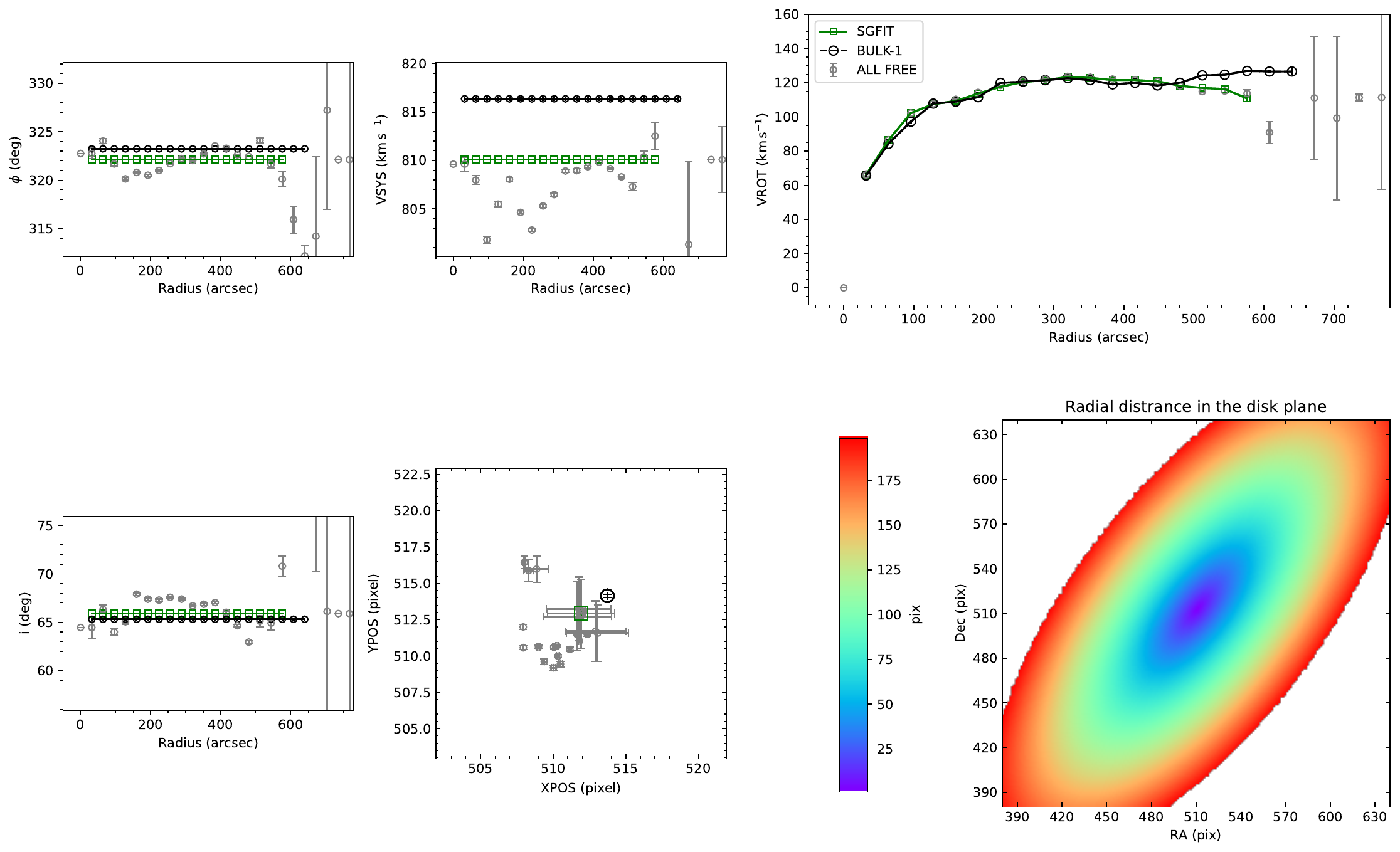}
    \vspace{-5mm} % Adjust the value as needed
    \caption{Initial 2D tilted-ring analysis of the single Gaussian fitting velocity field (SGFIT, open squares) and the first extracted bulk velocity field (BULK-1, filled and open black circles) for $\phi$ (position angle), VSYS (systemic velocity), $i$ (inclination), XPOS and YPOS (kinematic centre), and rotation velocity (VROT). Gray open circles represent the fit results with all ring parametres free (ALL FREE). The corresponding radial distances in the disk plane from the kinematic centre to the pixels of the 2D map are shown in the lower-rightmost panel. See Section~\ref{sec:2.3.5} for details.}
    \label{fig:3}
\end{figure*}

We conduct an initial 2D tilted-ring analysis using the velocity field derived from either single Gaussian (or Hermite Gaussian) fits or {\sc moment1} maps. This analysis provides the initial estimates of the 2D kinematical ring parametres of the galaxy's disk, such as kinemtic centre (XPOS, YPOS), position angle (PA), and inclination (INCL) which are then used for deriving the radial distances of each LOS pixel in the cube from the kinematic centre. In this work, we use {\sc 2dbat-PI}\footnote{https://github.com/seheon-oh/2dbat-PI} which fits a non-parametric 2D tilted-ring model with B-spline models for the ring parametres to a velocity field. For NGC 4559, we adopt constant parametres for PA and INCL as well as the kinematic cenre (XPOS, YPOS) and systemic velocity (VSYS). The derived ring parametres including the rotation velocity (VROT) from the {\sc 2dbat-PI} analysis are presented in Fig.~\ref{fig:3}. The resulting radial distances of individual LOS pixels of the cube which are derived using the ring parametres are mapped in the panel (d) of Fig.~\ref{fig:3}. 

\subsection{Selection criteria for bulk motion gas}\label{sec:2.3}

\subsubsection{Radial profile of \hi surface density}\label{sec:2.3.1}

\hi surface density profiles in disk galaxies often exhibit radial variation due to features in the disk like spiral arms and bars. Generally, these profiles flatten or decline near the centre and extend beyond the stellar disk in the outer regions, where they typically follow an exponentially declining radial distribution. \citet{2014MNRAS.441.2159W} found that the outer regions show a homogeneous radial distribution of \hi, well-fit by an exponential function with a scale length of 0.18 \(R_{1}\). This was determined using the azimuthally averaged radial profiles of \hi gas in 42 galaxies from the Bluedisk survey, which are robustly reproduced by both smoothed-particle hydrodynamics (SPH) simulations and semi-analytic models (\citealt{2014MNRAS.441.2159W}). To describe the shape of a two-component \hi radial profile-constant or depressed at the centre and exponential in the outer region—and to obtain its deconvolved shape, \citet{2014MNRAS.441.2159W} adopted an analytic expression of the form (see also \citealt{2002A&A...390..829S}; \citealt{2012ApJ...756..183B}):

\begin{equation}\label{eq:1}
\Sigma(r) = \frac{I_{1} \, \exp\left(-\frac{r}{r_{s}}\right)}{1 + I_{2} \, \exp\left(-\frac{r}{r_{c}}\right)},
\end{equation}

\noindent where \( I_{1} \), \( I_{2} \), \( r_{s} \), and \( r_{c} \) are free parametres. According to \cite{2014MNRAS.441.2159W} as $r$ increases, Eq.~\ref{eq:1} simplifies to an exponential profile with a scale radius of $r_{\rm s}$. The parameter $r_{c}$ represents the characteristic radius where the profile flattens in the inner region. As noted by \cite{2012ApJ...756..183B}, the $\rm HI + H_{2}$ gas exhibits an exponential radial distribution. The parameter $I_{1}$ corresponds to the $\rm HI + H_{2}$ gas, while $I_{2}$ pertains to the $\rm HI / H_{2}$ ratio at the centre.

In the first row panels of Fig.~\ref{fig:4}, we present the integrated intensities of all the optimally decomposed Gaussian components plotted against their distances \( r \) from the kinematic centre (XPOS, YPOS) in the galaxy plane. The radial distances \( r \) are derived using the geometrical ring parametres obtained from the initial 2D tilted-ring analysis—kinematic centre (XPOS, YPOS), position angle (\(\phi_{\text{init}}\)), and inclination (\(i_{\text{init}}\))—as described in Section~\ref{sec:2.2}. The dashed line indicates the fit of the model radial \hi surface density profile, given by Eq.~\ref{eq:1}, to the integrated intensities. Despite some scatter, the radial distribution of the \hi surface density of the decomposed Gaussian components in NGC 4559 is consistent with the template model profile. Given the uncertainties in the initial tilted-ring parametres, which result in uncertainties in \( r \), we adopt a 3-sigma boundary when selecting candidate Gaussian components for bulk motions. This boundary can be further refined as more accurate ring parametres are obtained in subsequent iterations. This serves as the first necessary condition for extracting bulk \hi gas motions in the disk of NGC 4559.

\subsubsection{Radial profile of \hi velocity dispersion}\label{sec:2.3.2}

Observations of \hi gas velocity dispersion in galaxy disks show that the dispersion generally decreases exponentially with radius (e.g., \citealt{1996IAUS..169..489D, 2009AJ....137.4424T, 2015AJ....150...47I}). Late-type spiral galaxies often exhibit a clear radial decline in velocity dispersion, well-modelled by an exponential function, while dwarf galaxies often display flatter profiles. Several factors, such as stellar feedback, spiral arms, tidal interactions, and ram-pressure stripping by the intergalactic medium (IGM), can stir up the \hi gas in the disk, leading to increased velocity dispersion (\citealt{2009AJ....137.4424T}). \citet{2009AJ....137.4424T} (see also \citet{2013ApJ...765..136S}) indicate that \hi velocity dispersion of dwarf or spiral galaxies generally decreases with radius, from around 20 \kms\, near the centre to approximately 5 km/s in the outer regions.

Similar to the \hi surface densities of the decomposed Gaussian components, the second row panels of Fig.~\ref{fig:4} shows the radial (\(r_{\text{init}}\)) distribution of velocity dispersion for all the decomposed Gaussian components of NGC 4559. We model the radial velocity dispersion profiles using an exponential function of the form:

\begin{equation}\label{eq:2}
\sigma(r) = \sigma_0 \, \exp\left(-\frac{r}{r_\sigma}\right),
\end{equation}

\noindent where \( \sigma_0 \) represents the central velocity dispersion and \( r_\sigma \) is the scale radius. It is noteworthy that the exponential fit results can vary depending on the radial range considered. \citet{2015AJ....150...47I} showed that in some galaxies, the radial decline in velocity dispersion starts to flatten near the optical radius. Therefore, fitting only within \( R_{25} \) would produce a steeper slope. In our analysis, we fit the velocity dispersion, ranging from 5\,\kms\, (approximately the channel resolution) to 20\,\kms, within 2\( R_{25} \). In the first row panels of Fig.~\ref{fig:4}, the exponential fit as shown by the dashed line. As with the \hi surface density criteria in Section~\ref{sec:2.3.1}, we use a 3-$\sigma$ boundary for the \hi velocity dispersion criteria, i.e., the shaded region in Fig.~\ref{fig:4}, for selecting bulk motion gas candidates. Additionally, we ensure that the kinetic energy criterion selects pixels that satisfy both the column density and velocity dispersion criteria. This approach guarantees that only pixels with valid column densities and velocity dispersions are included in the final selection for bulk motion gas, as discussed in the following section.

\subsubsection{Radial profile of \hi kinetic energy}\label{sec:2.3.3}

The kinetic energy of H{\sc i} gas clouds in a galaxy's disk is directly related to the velocity dispersion of the H{\sc i} gas, being associated with the dynamics and stability of the gas disk. Both thermal broadening and turbulence contribute to the overall \hi gas velocity dispersion in the disk of galaxies, which typically range from 5 to 15 km/s on small scales \citep{2009AJ....137.4424T}. This \hi gas velocity dispersion, essential for preventing gravitational collapse of the gas, reflects the combined effects of various energy sources, including star formation feedback and magnetorotational instabilities (MRI) (e.g., \citealt{2014MNRAS.438.1552F}). \citet{2009AJ....137.4424T} found a systematic radial decline in \hi line width across 11 disk galaxies from THINGS\footnote{The \hi Nearby Galaxy Survey (\citealt{2008AJ....136.2563W})}, suggesting a corresponding decrease in kinetic energy density with radius. Inside the galactocentric radius \( R_{25} \), \hi velocity dispersion remains significantly above thermal values, implying turbulence driven by stellar feedback like supernova explosions, with a positive correlation observed between \hi kinetic energy and star formation rate (see \citealt{2009AJ....137.4424T}).

The radial profile of kinetic energy for \hi gas in a galaxy's disk can be modelled based on the exponential decline of both \hi gas velocity dispersion and surface density. Let the velocity dispersion and surface density be described by Eqs.~\ref{eq:1} and \ref{eq:2}, respectively. The kinetic energy per unit area, expressed as \(E_k(r) = 1.5 \times \Sigma(r)\ \times \sigma(r)^{2} \), thus follows an exponential decay. The factor \(1.5\) accounts for all three velocity components, assuming isotropic velocity dispersion as described in \citet{2009AJ....137.4424T}. Substituting the radial profiles for \(\sigma(r)\) and \(\Sigma(r)\), we obtain:

\begin{equation}\label{eq:3}
E(k) = \frac{3}{2} \, I_1 \, \sigma_0^2 \, \frac{\exp\left(-\frac{r}{r_s} - 2 \, \frac{r}{r_{\sigma}}\right)}{1 + I_2 \, \exp\left(-\frac{r}{r_c}\right)}
\end{equation}

\noindent where \( I_{1} \), \( I_{2} \), \( r_{s} \), \( r_{c} \), \( \sigma_{0} \), and \( r_{\sigma} \) are the fitted parametres in Eqs.~\ref{eq:1} and \ref{eq:2}. This indicates that the kinetic energy density declines exponentially with radius, with a combined scale radius of the individual scale radii of velocity dispersion and surface density.

In the third row panels of Fig.~\ref{fig:4}, we plot the \hi kinetic energy per unit area for all the decomposed Gaussian components against their galactocentric distance normalized by \( R_{25} \). The dashed line in the figure shows the model radial profile of kinetic energy, constructed using the model radial profiles of \hi gas surface density and velocity dispersion in the first and second raw panels of Figs.~\ref{fig:4}, respectively. On average, the model profile is consistent with the derived kinetic energy of the Gaussian components. This supplementary selection criterion for kinetic energy, based on the \hi gas surface density and velocity dispersion, ensures that the kinetic energy of candidate Gaussian components for bulk motions is consistent with that of \hi gas characterised by the model \hi surface density and velocity dispersion.

\subsubsection{Rotation curve model}\label{sec:2.3.4}

\hi gas clouds consistent with the average values of velocity dispersion, surface density, and kinetic energy of bulk \hi gas following the circularly rotating gas disk may exhibit deviations from the underlying global kinematics of the galaxy. These deviations can manifest against the average disk rotation, such as high-velocity \hi gas clouds (\citealt{2018MNRAS.474..289W}), extraplanar lagging \hi gas (\citealt{2002AJ....123.3124F, 2005A&A...439..947B, 2019A&A...631A..50M}), or even counter-rotating anomalous \hi gas clouds (\citealt{2009ApJ...699...76H}). 

The radial, line-of-sight velocity, \( v_{\mathrm{LOS}} \), at rectangular sky coordinates \( (x, y) \) or pixel \( (x, y) \) is given as follows for a distant galaxy, where the flat-sky assumption can be applied (\citealt{1974ApJ...193..309R, 1989A&A...223...47B}):

\begin{equation}\label{eq:4}
v_{\mathrm{LOS}}(x, y) = v_{\mathrm{SYS}} + v_{\mathrm{ROT}} \, \cos(\theta) \, \sin(i) + v_{\mathrm{EXP}} \, \sin(\theta) \, \sin(i),
\end{equation}

\noindent where \( v_{\mathrm{SYS}} \), \( v_{\mathrm{ROT}} \), \( v_{\mathrm{EXP}} \), \( i \), and \( \theta \) are the systemic velocity, rotation velocity, expansion velocity, inclination, and azimuthal distance from the major axis, measured counter-clockwise in the plane of the galaxy, as derived from the tilted-ring analysis in Section~\ref{sec:2.2}, and \( \cos(\theta) \) is:

\begin{equation}\label{eq:5}
\cos(\theta) = \frac{-(x - \text{XPOS}) \, \sin(\phi) + (y - \text{YPOS}) \, \cos(\phi)}{r}.
\end{equation}

In this work, assuming VEXP is zero (i.e., no expansion velocities in the disk kinematics) and deriving the radial distance \( r \) from the kinematic centre (XPOS, YPOS) in the disk plane, as well as the inclination \( i \) and position angle \( \phi \) measured in the plane of the sky at rectangular sky coordinates \( (x, y) \), we calculate the rotation velocities \( v_{\mathrm{ROT}} \) of the decomposed \hi Gaussian components in Section~\ref{sec:2.1} at \( (x, y) \) as follows:

\begin{equation}\label{eq:6}
v_{\mathrm{ROT}}(x, y) = \frac{v(x, y) - v_{\mathrm{SYS}}} {\cos \theta(x, y) \, \sin i(x, y)}.
\end{equation}

The fourth row panels of Fig.~\ref{fig:3} show the rotation velocities of all the decomposed Gaussian components with radius in the galaxy plane, color-coded by their \hi surface density, velocity dispersion, and kinetic energy values. As shown, unlike the rotation velocity from the tilted-ring analysis, which provides ring-by-ring mean rotation velocity values, the rotation velocities of individual Gaussian components exhibit large scatter, though a dominant fraction of them likely follows the typical pattern of galaxy rotation curves, increasing with galaxy radius and reaching a maximum in the outer part, resulting in a flat rotation curve. The deviating \hi gas clouds could be associated with the stellar feedback.

In this paper, to provide an additional kinematic criterion for bulk \hi gas motions, which filters out any gas clouds significantly deviating from the global shape of the galaxy rotation curve, we adopt an observationally motivated empirical shape of galaxy rotation curves, the so-called 'universal rotation curve,' described in \citet{1996MNRAS.281...27P}. Following \citet{2008ApJ...684.1018G}, the universal rotation curve, derived from the rotation curve analysis of 50 spiral galaxies with respect to their luminosities, is given by:

\begin{equation}\label{eq:7}
v_{\mathrm{ROT}}(r) = v_{\mathrm{opt}} \, \sqrt{v_{\mathrm{disk}}^2(r) + v_{\mathrm{halo}}^2(r)},
\end{equation}

\noindent where \(v_{\text{opt}}\) is given at \(r_{\text{opt}}\) as,

\begin{equation}\label{eq:8}
v_{\mathrm{opt}} = \frac{200 \, \lambda^{0.41}}{\sqrt{0.80 + 0.49 \log_{10}(\lambda) + \frac{0.75 \exp(-0.4 \lambda)}{0.47 + 2.25 \lambda^{0.4}}}},
\end{equation}

\noindent and \(r_{\text{opt}}\) is defined as

\begin{equation}\label{eq:9}
r_{\mathrm{opt}} = 13 \, \lambda^{0.5}\, \mathrm{kpc},
\end{equation}

\noindent where \(\lambda\) is the luminosity ratio \(L/L^*\). Here, \(L^*\) represents the total blue luminosity of the galaxy, with \(\log L^* = 10.4\) in solar units \citep{2008ApJ...684.1018G}.

The disk velocity is given by:

\begin{equation}\label{eq:10}
v_{\mathrm{disk}}(r) = \sqrt{\left(0.72 + 0.44 \log_{10}(\lambda)\right) \, \frac{1.97 \, \left(\frac{r}{r_{\mathrm{opt}}}\right)^{1.22}}{\left(\frac{r}{r_{\mathrm{opt}}}\right)^2 + 0.61^{1.43}}}
\end{equation}

\noindent and the halo velocity is given by:

\begin{equation}\label{eq:11}
v_{\mathrm{halo}}(r) = \sqrt{1.6 \, \exp(-0.4 \lambda) \, \frac{\left(\frac{r}{r_{\mathrm{opt}}}\right)^2}{\left(\frac{r}{r_{\mathrm{opt}}}\right)^2 + 2.25 \lambda^{0.4}}}.
\end{equation}

\noindent The rotation velocity is thus obtained by combining the disk and halo components, scaled by \( v_{\mathrm{opt}} \).

\begin{figure*}
    \hspace{-5mm} % Increase the negative value to move further left
    \includegraphics[width=1.0\textwidth]{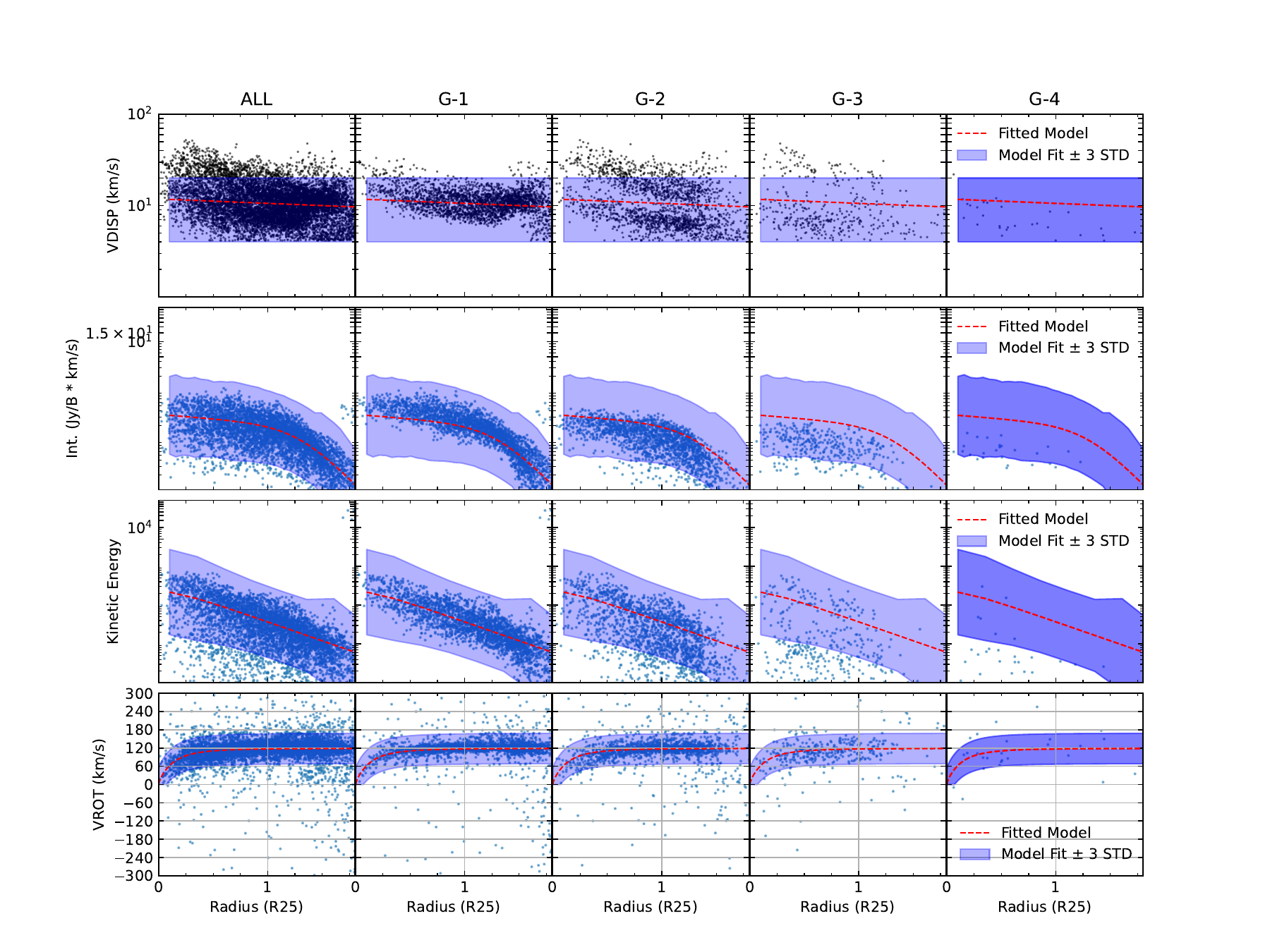}
    \vspace{0mm} % Adjust the value as needed
    \caption{Radial profiles of integrated intensities (Int.), velocity dispersions (VDISP), kinetic energies, and rotation velocities for all decomposed Gaussian components of NGC 4559's \hi data cube, shown from top to bottom rows (ALL), along with those for each individual Gaussian component (G1, G2, G3, and G4) from the profile decomposition analysis in Section~\ref{sec:2.1}. The ring parametres (XPOS, YPOS, VSYS, $\phi$, and $i$) are derived from the 2D tilted-ring analysis of the initially extracted bulk velocity field (BULK-1), which are used to calculate the radial distance from the kinematic centre in the disk plane and the rotation velocity of the individual Gaussian components. The radial distance is given in units of \(R_{25}\). The dashed lines represent model fits to the radial profiles, and the shaded regions indicate the $\pm$3-$\sigma$ boundaries as described in Section~\ref{sec:2.3}.}
    \label{fig:4}
\end{figure*}

While some cases of galaxy rotation curve analysis do not fit the proposed universal rotation curve (e.g., \citealt{1999ASPC..182..339B}; \citealt{1997PhDT........13V}), the universal rotation curve provides a reasonable description for galaxy rotation curves and serves as a useful observational basis for the circular rotation components of \hi in the disk. We fit the universal rotation curve given in Eq.~\ref{eq:7} to the rotation velocities of all the decomposed Gaussian components, with \( r_{\text{opt}} \) and \( \lambda \) as free parametres. The best-fit result is overplotted as the dashed line in the VROT panels of Fig.~\ref{fig:4}. Conservatively, we classify the Gaussian components whose rotation velocities fall outside the 3-$\sigma$ upper and lower or negative velocity boundaries with respect to the best-fit universal rotation curve as potential candidates for high-velocity clouds or lagging gas clouds, which could be associated with stellar feedback or observational m  effects, particularly in the inner region.

\subsubsection{2D tilted-ring analysis on the velocity field of bulk motion gas candidates}\label{sec:2.3.5}

\begin{figure*}
    \includegraphics[width=1.0\textwidth]{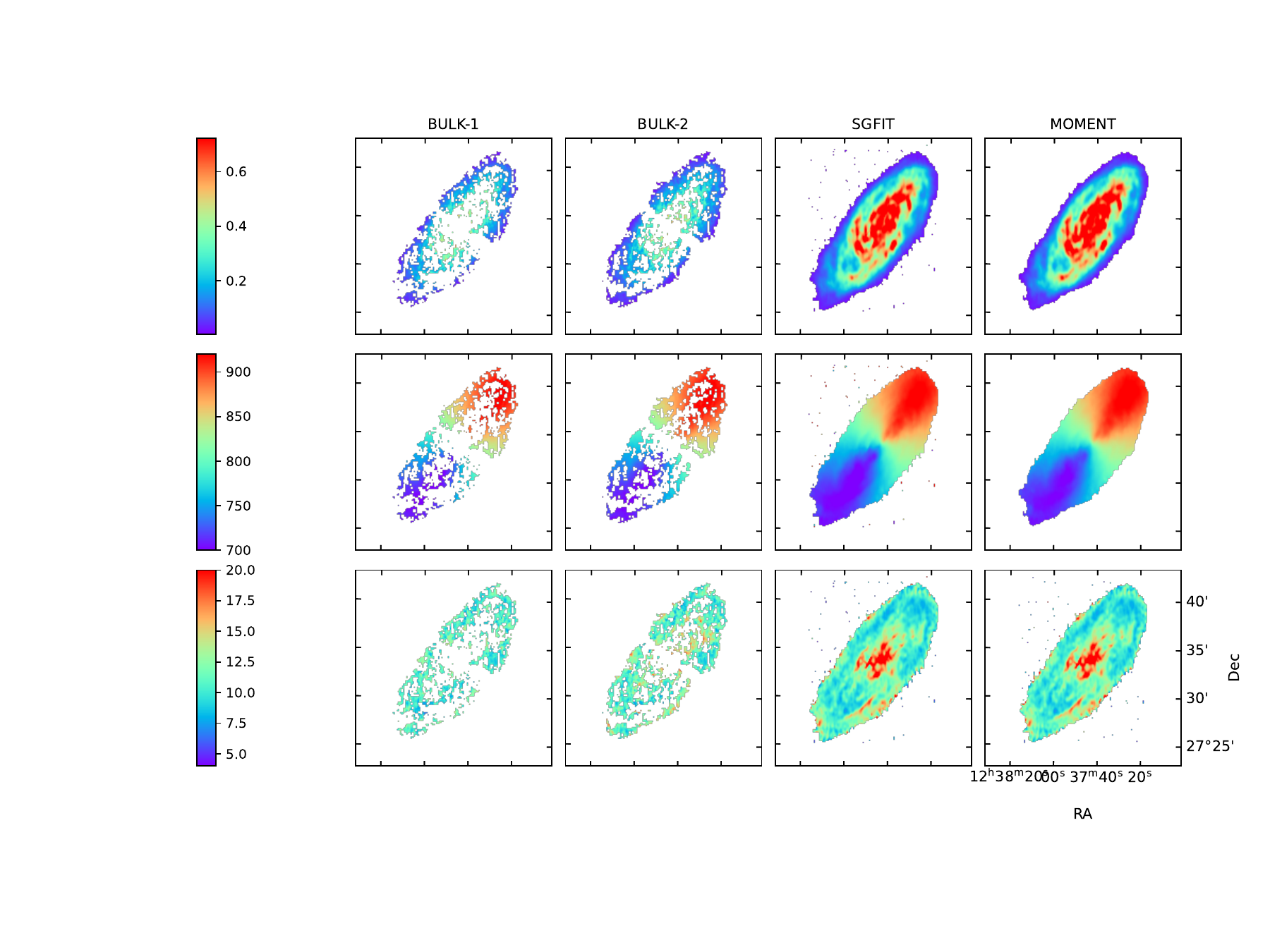}
    \hspace{0mm}
    \vspace{-20mm} % Adjust the value as needed
    \caption{2D maps of the first and second extracted bulk velocity fields (BULK-1 and BULK-2), the single Gaussian fitting velocity field (SGFIT), and {\sc moment1} extracted from the \hi data cube of NGC 4559 from the WSRT-HALOGAS survey, showing: (1) integrated intensity (Int.), (2) line-of-sight velocity (VLOS), and (3) velocity dispersion (VDISP).}
    \label{fig:5}
\end{figure*}

The first extracted velocity field of the bulk \hi gas motion candidates, derived using the selection criteria based on (1) surface density, (2) velocity dispersion, (3) kinetic energy, and (4) rotation curve shape, is shown in the first column panels of Fig.~\ref{fig:5}. Regions that do not satisfy all the selection criteria or where the peak S/N of the Gaussian fit is less than 3 are left blank. It is important to note that this initial bulk velocity field is not only dependent on the adopted selection criteria but also on the tilted-ring parametres derived using the single Gaussian fit velocity field or moment1 map, which were used to compute the radial distance from the kinematic centre in the galaxy plane.

Therefore, we perform an additional 2D tilted-ring analysis using the initially extracted bulk velocity field to update the ring parametres (XPOS, YPOS, VSYS, PA, and INCL) and, consequently, the radial distance of all the decomposed Gaussian components in the galaxy plane. As described in Section~\ref{sec:2.2}, we apply {\sc 2dbat-PI} to the velocity field, assuming constant ring parametres except for \( V_{\text{ROT}} \), with cubic spline regularization; if necessary, higher-order fits for PA, INCL, and \( V_{\text{ROT}} \) can be applied. In Fig.~\ref{fig:3}, we compare the 2D tilted-ring parametres with galaxy radius derived using the single Gaussian fit velocity field and the initially extracted velocity field for the bulk motions. On average, the ring parametres over radii between the two cases are not significantly different, though some radial regions, such as the region from 5 to 7 kpc, show differences. As shown in the N-Gauss map of Fig.~\ref{fig:2}, this could be associated with higher non-Gaussianity, as quantified by large N-Gauss values. The single Gaussian fit velocity field may have been affected by these turbulent gas motions, leading to large uncertainties in determining the central velocities of the profiles in the corresponding region.

\subsection{Extraction of bulk and non-bulk \hi gas motions}\label{sec:2.4}
As schematized in Fig.~\ref{fig:1}, we return to the first step of the procedure, now using the updated radial distances of all the decomposed Gaussian components and the model velocity field derived from the initially extracted bulk velocity field. We then continue the procedure to derive secondary maps for the bulk motions. If necessary, we iterate these procedures until the consecutive 2D maps for the bulk motion gas candidates, such as the integrated intensity, velocity field, and velocity dispersion maps, converge. We calculate the difference between consecutive velocity fields by measuring the mean and root-mean-square (rms) values of the residual map. For each iteration, we compare the current mean value of the residual map (representing the velocity difference) with the value from the previous iteration. If the difference lies within \(\pm 3\sigma\) of the mean residual value, we consider the differences insignificant and stop the iterative procedure. If convergence is not achieved, we proceed to the next iteration until the threshold is met. For NGC 4559, we find that the mean differences between the secondary and tertiary 2D maps extracted for the bulk motions are insignificant, and therefore, we use the secondary maps as the final ones for the bulk motions.

In Fig.~\ref{fig:5}, we present the integrated intensity, velocity field, and velocity dispersion 2D maps of the candidate \hi bulk motion gas extracted through the iterations of the procedure described above. The final maps, from the second iteration for NGC 4559, are shown in the BULK-2 panels. Additionally, we also show the \hi column density maps of the bulk and non-bulk motion gases. Firstly, compared with the {\sc moment1} map, less \hi gas is classified as bulk motion in the outer regions, resulting in a less extended extracted bulk velocity field. As shown in the VDISP panels of Fig.~\ref{fig:4}, this is mainly due to the decomposed \hi gas with high velocity dispersion in the outer region. These \hi gas clouds with high velocity dispersion, did not pass the bulk selection criteria described in Section~\ref{sec:2.3}. The properties of these non-bulk \hi gas clouds are more likely to resemble those of diffuse \hi gas with low column density and high velocity dispersion. The high velocity dispersion of non-bulk \hi gas with low surface density in the outer regions could be associated with star formation processes in the galaxy or tidal interaction with others, which drive turbulent gas motions \citep{2005A&A...439..947B, 2017ApJ...839..118V}.

While star formation in the outer parts of NGC 4559 is limited, localized regions of star formation may still influence the gas kinematics. Additionally, external processes such as tidal interactions and radial gas inflows, as suggested by recent FEASTS data (\citealt{2023ApJ...944..102W}; Lin+ in prep), may also play a role. Recent deep \hi observations of the NGC 4631 group show that tidal interactions may induce CGM cooling and gas accretion through the production of diffuse \hi gas \citep{2023ApJ...944..102W}. CGM cooling can result in radial gas inflows into the host galaxy, bringing the system to a dynamically inequilibrium state. Given this, part of the non-bulk \hi gas may be associated with tidal interactions or the CGM. These perturbations, along with potential variations in inclination or position angle, could explain the residuals classified as non-bulk motions in our analysis. In this respect, deep \hi observations like FEASTS will be useful for investigating such non-bulk \hi gas in and around nearby galaxies.

The bulk velocity field shows a regular rotation pattern, and, on average, the velocity dispersion values of the bulk motion gas are lower than those in both the single Gaussian fit velocity dispersion map and {\sc moment2}, particularly in the inner region. The gas with high velocity dispersion in the inner regions could be associated with stellar feedback processes in the disk. However, this effect could also result from observational beam smearing in regions with steep velocity gradients, such as the inner parts of the galaxy. The remaining gas that does not pass at least one of the selection criteria for bulk motion are classified as non-bulk motion gas. We will discuss non-bulk motion gas in Section~\ref{sec:4}.

\begin{figure*}
    \includegraphics[width=1.0 \textwidth]{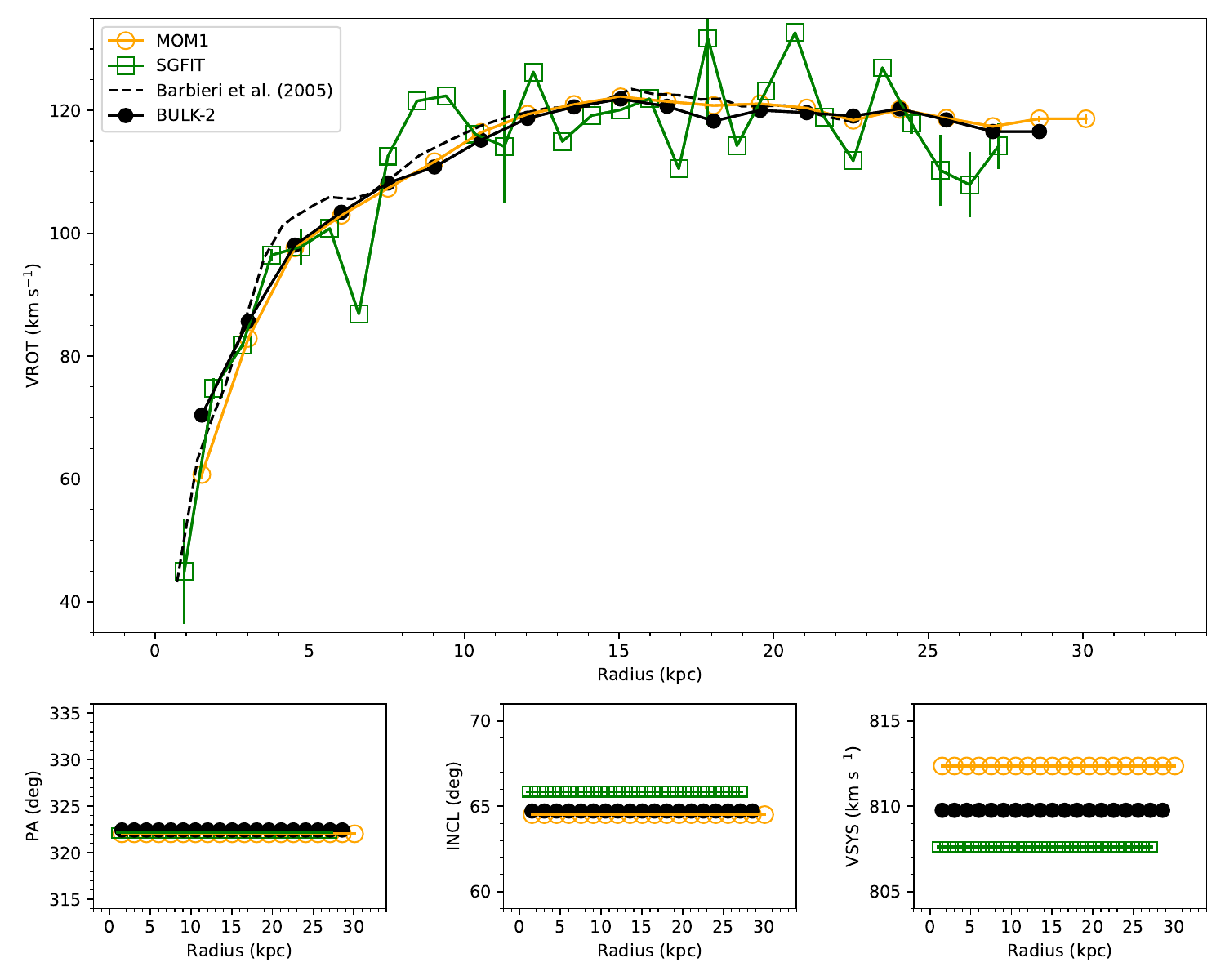}
    \vspace{-5mm} % Adjust the value as needed
    \caption{Upper panel: Rotation curves of NGC 4559. Black filled circles represent the best-fit results from the 2D tilted-ring analysis using the second bulk velocity field (BULK-2). Upward and downward triangles indicate the fit results for the receding and approaching sides of the bulk velocity field, respectively. Open circles and squares represent the rotation curves derived from the {\sc moment1} and single Gaussian fitting velocity field (SGFIT), respectively. Lower panels: Same as the upper panel, but for position angle ($\phi$), inclination ($i$), and systemic velocity (VSYS). See Section~\ref{sec:3} for more details.}
    \label{fig:6}
\end{figure*}

\begin{figure*}
    \includegraphics[width=1.0\textwidth]{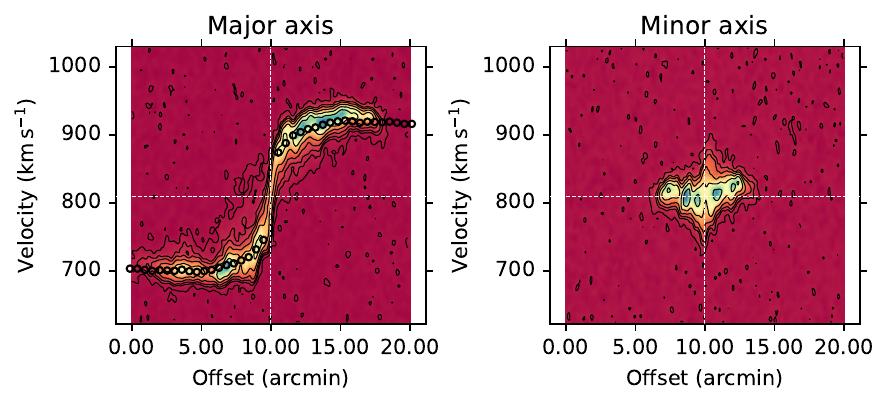}
    \vspace{-5mm} % Adjust the value as needed
    \caption{\hi position-velocity (position-velocity diagram at 28\arcsec\,resolution along the major and minor axes of NGC 4559 (\( \phi_{\rm major} = 322^{\circ} \), \( \phi_{\rm minor} = 232^{\circ} \), \( v_{\mathrm{SYS}} = 810 \) \kms). Contours are -2, 2, 4, 8, 16, 32, and 64$\sigma$ levels. The black open circles indicate the projected BULK-2 \hi gas rotation curve. The white dashed lines indicate the systemic velocity and kinematic center derived from the tilted-ring analysis in Section~\ref{sec:3}.}
    \label{fig:7}
\end{figure*}

\section{Rotation curve of \hi bulk motion gas in NGC 4559}\label{sec:3}

We derive the rotation curve of the \hi bulk motion gas in NGC 4559 using the bulk velocity field extracted in Section~\ref{sec:2.3}. For this purpose, we utilize {\sc 2dbat-PI}—a Python3 reimplementation of the original {\sc 2dbat} in C \citep{2018MNRAS.473.3256O}. In our analysis, we run {\sc 2dbat-PI} on the bulk velocity field with two sets of fits: a low-order fit (constant PA and INCL with a cubic spline model for VROT) and a high-order fit (cubic spline models for PA, INCL, and VROT). The resulting bulk rotation curves are presented in Fig.~\ref{fig:6}.

Previous studies, such as \citet{2005A&A...439..947B, 2017ApJ...839..118V}, have reported the presence of extra-planar \hi gas deviating from the circularly rotating gas disk. These deviations were identified by analyzing the residuals in the velocity field after subtracting the tilted-ring model velocity field from the moment1 map. While these extra-planar gas clouds generally follow the circular rotation of the disk, they are found to lag behind by up to 30 km/s compared to the dominant circular motion of the galaxy \citep{2005A&A...439..947B}. Similar extra-planar gas motions were identified in the WSRT-HALOGAS data cube for NGC 4559 using 3D tilted-ring analysis with TiRiFiC \citep{2017ApJ...839..118V}.

In Fig.~\ref{fig:6}, we compare the \hi bulk rotation curve derived in this work with the one presented in \citet{2005A&A...439..947B}, as well as with the ones derived from the {\sc moment1} map of the disk components. Notably, the \hi rotation curve derived using both sides of the bulk velocity field, extracted by tracing decomposed \hi gas with dominant \hi properties (surface density, velocity dispersion, kinetic energy) and modelled using the universal rotation curve in a statistically manner, is systematically higher in the inner region ($<$3 kpc) and lower in the intermediate region (approximately 3 to 20 kpc) than the one presented by \citet{2005A&A...439..947B}. This discrepancy could be influenced by the lagging gas components offset by up to 30 \kms as found in \citet{2005A&A...439..947B}. These non-bulk \hi gas components could introduce uncertainties in the measurements of representative line-of-sight velocities, deviating from the bulk motions. The inclination difference ($\sim$$1^{\circ}$) between our results and those of \citet{2005A&A...439..947B} is small, so the inclination effect is insignificant.

%--------------------------------------------------------------------------------------

\begin{figure*}
    \includegraphics[width=1.0\textwidth]{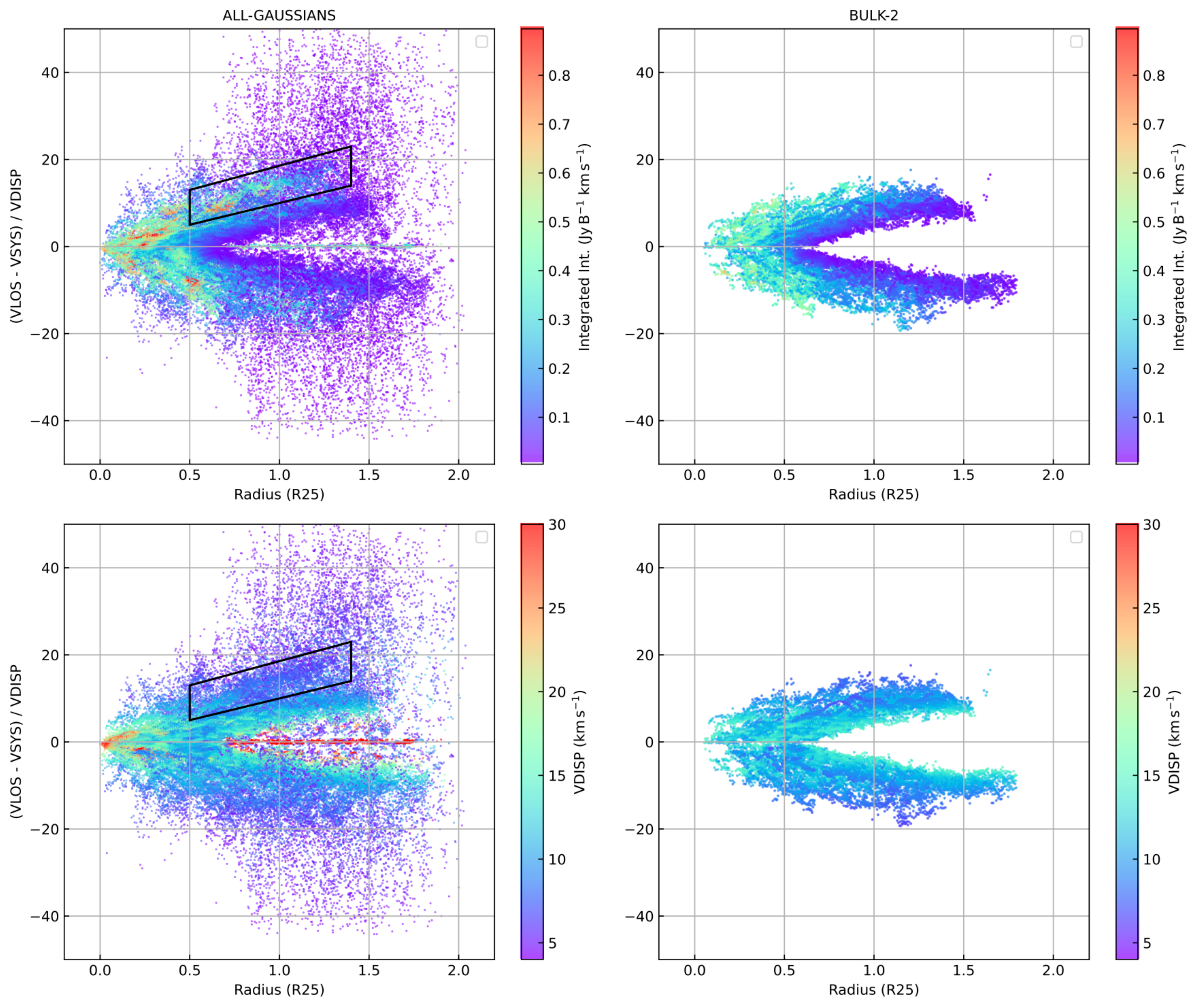}
    \vspace{-5mm} % Adjust the value as needed
    \caption{Projected phase-space diagram of all Gaussian components (ALL-GAUSSIANS) and the second extracted bulk motion gas (BULK-2). The ring parametres (XPOS, YPOS, VSYS, $\phi$, and $i$) are derived from the 2D tilted-ring analysis using the BULK-2 velocity field. These parameters are used to calculate the projected radial distance from the kinematic centre in the disk plane and the rotation velocity of individual Gaussian components, as described in Section~\ref{sec:3}. The projected radial distance is normalised by \(R_{25}\). The decomposed Gaussian components are color-coded based on their integrated intensities (upper panels) and velocity dispersions (lower panels), respectively. The upper split, indicating the non-bulk gas components that deviate from the expected disk plane kinematics, is enclosed within a quadrilateral. See Section~\ref{sec:4.2} for more details.}
    \label{fig:8}
\end{figure*}

\begin{figure*}
    \includegraphics[width=1.0\textwidth]{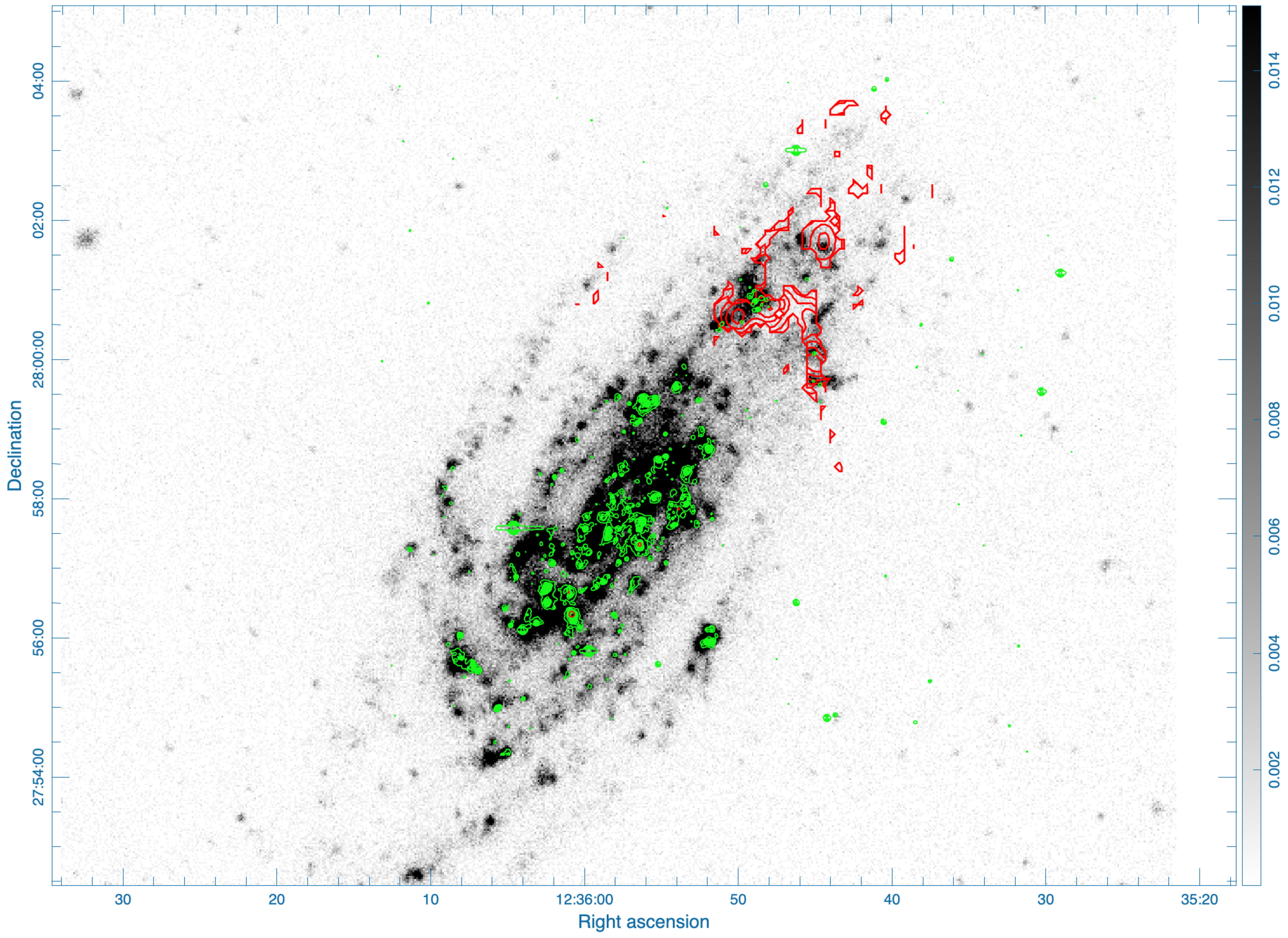}
    \vspace{-5mm} % Adjust the value as needed
    \caption{GALEX {\it FUV} image of NGC 4559 (in units of counts per second), overlaid with KPNO 2.1m {\it H$\alpha$} data (green contours). The non-bulk motion gas in the upper split, visually boxed in Fig.~\ref{fig:8}, is contoured in red, with intensity levels ranging from 0.06 to 1.0 in steps of 0.13 $\rm Jy\,beam^{-1}\,km\,s^{-1}$. See Section~\ref{sec:4.2} for more details.}
    \label{fig:9}
\end{figure*}

\section{Non-bulk \hi gas motions in NGC 4559}\label{sec:4}

According to the selection criteria for bulk motions adopted in this paper, a significant amount of \hi gas in the WSRT-HALOGAS \hi data cube of NGC 4559 are classified as non-bulk \hi gas. Approximately 50\% ($\sim$$3.43 \times 10^{9}$ \Msun) of the total \hi gas ($\sim$$6.76 \times 10^{9}$ \Msun) extracted from the data cube, with a signal-to-noise ratio (S/N) greater than 3, is classified as non-bulk motions. These components deviate from the radial disk models in at least one of the selection criteria: gas surface density, velocity dispersion, kinetic energy, and rotation velocity. The deviations from the \hi gas disk are evident despite our relatively wide boundaries for the selection criteria (i.e., 3-sigma upper and lower limits relative to the exponential models).

These non-bulk motion gases are likely associated with various hydrodynamic processes within the galaxy or could be caused by gravitational interactions with galaxies in the course of galaxy evolution. They could include cold gas accretion from the cosmic web, high-velocity clouds around the galaxy that might have been expelled by stellar feedback and are now falling back to the disk, as well as outflowing or inflowing gas related to stellar feedback processes within the galaxy, large-scale gas accretion, and tidal interaction processes. To examine both the distribution and kinematics of these non-bulk gas components, high-sensitivity \hi observations like FAST-FEASTS (\citealt{2023ApJ...944..102W}) would be helpful (Oh et al. in preparation).

We suspect that the large fraction ($\sim$50\%) of non-bulk gas in NGC 4559 likely includes a combination of physical processes and observational effects. Non-bulk gas is not restricted to extra-planar \hi gas; in the inner disk, it may include projected streaming motions or localized gas out of equilibrium due to stochastic star formation, while in the outer disk, tidal disturbances could contribute to velocity irregularities. Observational effects such as beam smearing and projection uncertainties also play a role, highlighting the need for caution when interpreting this fraction and comparing it with previous studies, such as \cite{2017ApJ...839..118V}, which report a smaller fraction of extra-planar \hi gas.

\subsection{Position-velocity diagram}\label{sec:4.1}

In Fig.~\ref{fig:7}, we present the position-velocity (PV) diagram of NGC 4559, extracted from the WSRT-HALOGAS \hi data cube along the kinematic major axis, as defined from the 2D tilted-ring analysis in Section~\ref{sec:2.3.4}. Forbidden \hi gas (or ‘beard’ gas) is clearly visible in the PV diagram, consistent with findings from previous studies \citep{2005A&A...439..947B, 2017ApJ...839..118V}. We also overlay the bulk rotation curve along the kinematic major axis within one beam size ($\sim$28\arcsec) using thick solid and thin dashed lines, respectively. The line-of-sight velocity offset of these non-bulk motion gases relative to the bulk motion gas is significant, with differences reaching up to 30 km/s.

At certain radii, such as at $\sim$8 arcmin, \hi gas components with low column density, which are more consistent with the global kinematics of the disk, might not have been extracted as bulk motion gas if the profile decomposition and bulk extraction methods described in this work were not applied. In these regions, the central velocities of \hi profiles derived using moment or single Gaussian fit analyses are biased towards the high-column density gas components, which deviate from the global kinematics as shown in the position-velocity diagram.

\subsection{Phase-space diagram}\label{sec:4.2}

In Fig.~\ref{fig:8}, we present the phase-space diagram of NGC 4559, where the $\rm (V_{LOS} - V_{SYS}) / V_{DISP}$ is plotted against the radial distance $R$ from the kinematic centre in the disk plane. In this diagram, $V_{LOS}$ represents the line-of-sight velocity, $V_{SYS}$ is the systemic velocity, $V_{DISP}$ is the velocity dispersion, and $R$ is the radial distance. This diagram allows us to investigate both bulk and non-bulk gas components, as classified in Section~\ref{sec:4}.

The right column of Fig.~\ref{fig:8} shows the case of the bulk gas component, extracted based on the selection criteria described in Section~\ref{sec:2}. Both the integrated intensities and velocity dispersions of the bulk motion gas exhibit a gradual decrease with increasing radial distance, indicating a relatively smooth distribution. In contrast, the non-bulk gas components display a more complex, non-uniform distribution, with regions of higher intensity and velocity dispersion scattered across the disk. In particular, non-bulk components with lower intensities and velocity dispersions are predominantly located in the outer regions of the galaxy.

Interestingly, two distinct branch-like structures are evident in the receding side of the phase-space diagram (i.e., $\rm (V_{LOS} - V_{SYS}) / V_{DISP} > 0$) in the "ALL-GAUSSIANS" panels (top-left and bottom-left). The upper split, visually enclosed within a quadrilateral in Fig.~\ref{fig:8}, suggests the presence of non-bulk gas components deviating from the expected disk plane kinematics in NGC 4559. To investigate the origins of these deviations, Fig.~\ref{fig:9} overlays the integrated intensities of the non-bulk gas motions on the GALEX\footnote{Galaxy Evolution Explorer, \cite{2005ApJ...619L...1M}} far-ultraviolet ({\it FUV}) image, with contours from the Kitt Peak National Observatory (KPNO) 2.1m telescope {\it H$\alpha$} observations (yellow) added. These distributions show that the spatial locations of the non-bulk gas motions are broadly consistent with star-forming regions, as traced by {\it FUV} and {\it H$\alpha$} emissions. This spatial overlap suggests that the systematic deviations may be partially driven by stellar feedback, where energy injected by young stars impacts the surrounding interstellar medium.

In contrast, these branch-like features are less pronounced in the panels showing bulk motion gas. The bulk motion gases on both the receding and approaching sides shows a symmetric distribution, except in the outer regions ($R > 125$ pixels). Additionally, systematic gradients in both intensity and velocity dispersion are observed across the disk: intensity decreases towards the zero line ($\rm (V_{LOS} - V_{SYS}) / V_{DISP} = 0$), while velocity dispersion increases. This trend could be a consequence of the relatively broad selection boundaries we adopted ($\pm 3\sigma$) for velocity dispersion, surface density, kinetic energy, and rotation velocities, as described in Section~\ref{sec:2}. These boundaries may have allowed for the inclusion of gas components that are mildly perturbed, thus resulting in the observed gradients. A more refined approach, possibly using physically motivated boundaries derived from models or statistical analysis of galaxy samples, could help isolate the undisturbed bulk motion gas that better represents the disk's kinematics.

\subsection{Characterising non-bulk motion gas with a 3D approach}

There has been long-lasting efforts in the literature to separate and characterize the kinematically peculiar gas, corresponding to non-bulk motion gas classified in this study. Many latest developments utilize tilted ring models to fit the motions in the azimuthal, radial and vertical directions, allowing additional freedom of disk thickness and warps to account for complex flux distribution (e.g. \citealt{2019A&A...631A..50M}, \citealt{2023MNRAS.520..147L}). An idealized tilted-ring model of such, would enable physical modeling of the inflow and outflow of gas, two key processes driving evolution of galaxies, which is the primary bottleneck encountered by modern galaxy formation models (\citealt{2023ARA&A..61..473C}). However, because of the high degeneracy between geometry and velocity projections, compromises often have to be taken in order to obtain a meaningful model in the global statistical sense.

For example, thin disks are masked to prevent them from dominating the fitting weight (\citealt{2023MNRAS.520..147L}), or functional forms are assumed to reduce the number of free parameters (\citealt{2019A&A...631A..50M}). An additional limitation is that the intrinsically assumed azimuthal uniformity in tilted-ring models is reasonable for circular motions, but may be less so for non-bulk motions. After all, disk dynamics drive spiral arm and bar structures (\citealt{2016A&A...594A..86R}), and effective stellar feedbacks are often launched from massive stellar clusters (\citealt{2024ARA&A..62..369S}). Moreover, while the bulk motion gas may be well confined to the local gravity, and non-bulk motion gas can be more susceptable to external environmental perturbations like tidal forces and ram pressure (\citealt{2022A&ARv..30....3B}).

The methods and results presented in this paper, demonstrates an alternative way approaching the peculiar gas characterization.  We seem to also use parameterized models, but they are only used as reference frames to separate the gas into bulk and non-bulk ones. So far, we only use tilted ring models to derive the rotation curves with the bulk-motion gas component. So that, we are allowed to see the complex full 3D distribution of the non-bulk motion gas. This work, for the first time, reveals that:

1) The non-bulk motion gas has a wide distribution of \( v_{\mathrm{LOS}} / \sigma_v \) at a given radius, providing potentially valuable physical information that is otherwise lost in tilted ring models. It not only shows high velocity dispersions, as often discussed in previous studies (e.g., the extraplanar \hi, \citealt{2017ASSL..430..323F}), but also includes a significant low-velocity dispersion component. The non-bulk motions may include in-plane streaming motions driven by torques from non-axisymmetric structures or tidal forces, localized inflows and outflows associated with star formation and feedback activities, and shocks induced by these processes.

2) Identifying non-bulk motion gas displayed as branch structures in the projected phase-space diagram, and investigating their connection to the extraplanar HI in the future, may help justify or improve the previous operation of masking thin disks when fitting extraplanar gas motions (\citealt{2019A&A...631A..50M}). We acknowledge that potential contamination from beam smearing should be more accurately quantified and mitigated in future studies to enhance the scientific merit and potential of these findings.

In summary, a unique contribution of this method is its ability to provide the full 3D distribution of non-bulk motion gas, along with more precisely confined rotation curves. The first aspect is important because it preserves the azimuthal and radial complexities within the data, enabling more sophisticated physical modeling of gas states and the effects of inflows and outflows. The second aspect is also important, as it facilitates better quantification of the gravitational potential, which serves as the critical foundation for most physical processes to occur.

\section{Conclusion}\label{sec:5}
In this paper, we develop and apply a new method for extracting bulk \hi gas motions in galaxy disks, with a practical application to the spiral galaxy NGC 4559. Our approach improves upon traditional techniques like moment analysis and line profile fitting by employing a decomposition of the line-of-sight velocity profiles into multiple Gaussian components. These components are subsequently classified into bulk and non-bulk motion gases based on a set of criteria including \hi surface density, velocity dispersion, kinetic energy, and rotation velocity.

Our method includes an iterative refinement process through a 2D tilted-ring analysis, which updates the kinematic parametres of the galaxy disk as more accurate bulk motion maps are generated. This iterative approach ensures that the final bulk velocity field is as accurate as possible, with any remaining non-bulk motion gases clearly identified and excluded from the bulk motion analysis. This methodology enables a more robust identification of the gas components dominating the galaxy’s bulk rotation, while filtering out non-bulk motions that may result from baryonic processes such as stellar feedback.

The effectiveness of this method was demonstrated using the \hi data cube from the WSRT-HALOGAS. Through our analysis, we found that approximately 50\% of the \hi gas in NGC 4559 is classified as non-bulk motion gas. The non-bulk motion gas identified in this work exhibits distinct characteristics: it shows higher velocity dispersion in the inner region and significant deviations from the rotation velocities predicted by the universal rotation curves for spiral galaxies. As discussed in previous studies, the non-bulk motion gas may be influenced by internal processes such as stellar feedback, which can drive turbulence and cause deviations from the regular rotational pattern of the disk. In some cases, it could also result from external processes, such as gas accretion or interactions with the galaxy’s environment.

This method not only enables us to extract bulk motion gases and, consequently, reliable rotation curves, but also maximizes the utility of the rich kinematic information in the \hi data cube of a galaxy, which can be used to examine hydrodynamic processes in and around the galaxy disk. This approach would be useful for analyzing data cubes from high-sensitivity \hi observations, such as MeerKAT-MHONGOOSE (\citealt{2024A&A...688A.109D}) and FAST-FEASTS (\citealt{2023ApJ...944..102W}), by identifying diffuse \hi gas around galaxies in group or cluster environments that deviates from their bulk motion gases, and probing its kinematic properties.

%%%%%%%%%%%%%%%%%% Acknowledgement %%%%%%%%%%%%%%%%%%

\section*{Acknowledgements}
SHOH acknowledges a support from the National Research Foundation of Korea (NRF) grant funded by the Korea government (Ministry of Science and ICT: MSIT) (No. RS-2022-00197685). JW thank support of research grants from  Ministry of Science and Technology of the People's Republic of China (NO. 2022YFA1602902), National Science Foundation of China (NO. 12073002, 12233001, 8200906879), and the China Manned Space Project.

%%%%%%%%%%%%%%%%%% Data Availability %%%%%%%%%%%%%%%%%%

\section*{Data availability}
The \hi data cube and moment maps of NGC 4559 used in this work were downloaded from the WSRT-HALOGAS (\citealt{2011A&A...526A.118H}) webpage at https://www.astron.nl/halogas/data.php.



\bibliographystyle{mnras}
\bibliography{main} % if your bibtex file is called example.bib

\end{document}